\documentclass[11pt]{article}
\pdfoutput=1 

\usepackage{jheppub} 
                     
\usepackage{amssymb,amsmath,amsthm,graphicx}
\usepackage{latexsym}
\usepackage{bm}
\usepackage[]{hyperref}
\usepackage{cleveref}

\def \be {\begin{equation}}
\def \ee {\end{equation}}
\def \bea {\begin{eqnarray}}
\def \eea {\end{eqnarray}}

\def \a {\alpha}

\pdfoutput=1

\def\ie{{\it i.e.}}

\begin{document}

\title{Further evidence for the weak gravity - cosmic censorship connection}

\author[\sharp]{Gary~T.~Horowitz,}
\affiliation[\sharp]{Department of Physics, University of California, Santa Barbara, CA 93106, US} 

\author[\diamond]{Jorge~E.~Santos}
\affiliation[\diamond]{Department of Applied Mathematics and Theoretical Physics, University of Cambridge, Wilberforce Road, Cambridge CB3 0WA, UK} 

\emailAdd{horowitz@ucsb.edu}
\emailAdd{jss55@cam.ac.uk}
\abstract{
We have recently shown that a class of counterexamples to (weak) cosmic censorship in anti-de Sitter spacetime is removed if the weak gravity conjecture holds. Surprisingly, the minimum value of the charge to mass ratio necessary to preserve cosmic censorship is precisely the weak gravity bound. To further explore this mysterious connection, we investigate two generalizations: adding a dilaton or an additional Maxwell field. Analogous counterexamples to cosmic censorship are found in these theories if there is no charged matter. Even though the weak gravity bound is modified, we show that in each case it is sufficient to remove these counterexamples. In most cases it is also necessary.
}
\maketitle
\section{Introduction and Summary}
A class of counterexamples to weak cosmic censorship in anti-de Sitter (AdS) spacetime  was proposed in \cite{Horowitz:2016ezu} and studied numerically in \cite{Crisford:2017zpi}. They involve solutions where an electric field grows in time without bound, producing arbitrarily large curvature that is visible to infinity. It was suggested \cite{Vafa} that these counterexamples might be removed if the weak gravity conjecture \cite{ArkaniHamed:2006dz} holds, which states that any consistent quantum theory of gravity must have a stable particle with $q/m \ge 1$. The idea behind this suggestion is that if charged particles are present, they will be pair created in the growing electric field, and their backreaction might stabilize the electric field.

This possible connection between weak gravity and cosmic censorship was explored in 
\cite{Crisford:2017gsb}. Rather than tackle the difficult quantum field theory in curved spacetime problem mentioned above, the charged matter was modeled by a classical charged scalar field\footnote{This is a reasonable classical version of the weak gravity conjecture if the predicted particle has spin zero. We will assume this is the case.}. It was found that if the charge to mass ratio of the scalar field was large enough, the original Einstein-Maxwell solutions became unstable to turning on the scalar, and the electric field did not diverge. Remarkably, the minimum value of this ratio required to preserve cosmic censorship was found to be precisely the weak gravity bound.

To further explore this mysterious connection, we now investigate two generalizations. The first is to add a dilaton, \ie, a neutral scalar field $\phi$ with a coupling to the Maxwell field of the form $e^{-2\alpha \phi}F^2$ for some constant $\alpha$. In the asymptotically flat case, static charged black holes to this theory were found in \cite{Gibbons:1987ps,Garfinkle:1990qj}.  These black holes have an extremal limit with $Q^2/M^2 = 1+\alpha^2 $. The fact that extremal black holes have $Q^2 > M^2$  is a consequence of the result that (like  Reissner-Nordstrom black holes) there is no force between them, so the electrostatic repulsion must now balance both the gravitational and scalar attraction. The weak gravity bound in the presence of dilatons has been discussed in the literature (see, \emph{e.g.}, \cite{Heidenreich:2015nta}).  The new bound is simply that $q/m $ should violate the extremality bound for black holes: $q^2/m^2 \ge 1+\alpha^2 $.

We will first show that there are analogs of our counterexamples to cosmic censorship if we add a dilaton with any coupling $\alpha$. We will then show that these counterexamples are all removed if we add a charged scalar field satisfying the modified weak gravity bound. This would not be true if we used the original weak gravity bound. (As discussed in \cite{Crisford:2017gsb} and reviewed below, the weak gravity bound is slightly different in AdS, and we use the AdS version.)  
When $\alpha < 1$ we will also show that the weak gravity bound is not only sufficient but also necessary to preserve cosmic censorship for this class of examples. For $\alpha > 1$, we have not been able to establish that the bound is necessary, and it may be possible to lower the charge to mass ratio slightly. It is interesting to note that spherical charged black holes also behave very differently depending on whether $\alpha<1$ or $\alpha>1$. For $\alpha<1$, as one approaches extremality, the Hawking temperature $T$  goes to zero, just like the familiar Reissner-Nordstrom black holes. However, for $\alpha =1$, $T$ approaches a constant and for $\alpha>1$, $T$ diverges. If the bound is not strictly necessary for $\alpha > 1$, it might be related to this unusual black hole behavior.

The second generalization that we will consider is to add a second Maxwell field. It is easy to see that the weak gravity bound must again be modified in this case. This bound is supposed to allow extremal black holes to decay. With two Maxwell fields, an extremal black hole satisfies $M^2 = Q_1^2 + Q_2^2$. Suppose it tries to decay into two particles, one with $q_1 = Q_1$ and the other with $q_2 = Q_2$. If $m_i = q_i$, then $M^2 = m_1^2 + m_2^2 < (m_1+m_2)^2$. So there is not enough energy in the black hole to create the two particles. The general condition on the charge to mass ratios of the particles that allows a black hole with multiple charges to decay was derived in  \cite{Cheung:2014vva}. This is the weak gravity bound that we will investigate.

We study a theory with two Maxwell fields and two charged scalars coupled to gravity (with a negative cosmological constant). Without the scalars, there are counterexamples to cosmic censorship as before. We then add the scalars and find that the new weak gravity bound is again precisely what is needed to  remove these counterexamples. This is highly nontrivial as the bound on $q_1/m_1$ depends on $q_2/m_2$ (and vice versa) and the two scalar fields interact with each other only through gravity.

We find these results quite mysterious. 
 The main open problem raised by this work is to understand why there is such a close connection between cosmic censorship and weak gravity, two conjectures that appear totally unrelated.

It is perhaps ironic that for many years people hoped that cosmic censorship would fail so that we had the possibility of observing effects of quantum gravity. Now we find that a conjecture about quantum gravity is preserving cosmic censorship. It appears that quantum gravity wants to remain hidden.

\section{Review of previous work} 
In this section we review the earlier work showing a connection between the cosmic censorship and  weak gravity conjectures. 
Consider  the bulk action
\begin{equation}
S = \frac{1}{16\pi G}\int \mathrm{d}^4 x\,\sqrt{-g}\,\Big(R+{6}-F^{ab}F_{ab}\Big)\,,
\label{eq:action0}
\end{equation}
 $F \equiv \mathrm{d} A$ is the Maxwell field and we have set the AdS length scale to one. With AdS boundary conditions, one is free to specify the (conformal) boundary metric at asymptotic infinity, as well as the asymptotic form of the vector potential $A_a$. We choose the boundary metric to be flat (as in standard Poincar\'e coordinates for AdS)  
\begin{equation}
\mathrm{d} s^2_\partial = -\mathrm{d} t^2+\mathrm{d} r^2+r^2 \mathrm{d}\varphi^2\;,
\label{eq:bndisflat}
\end{equation}
and the potential to asymptotically have only a nonzero time component of the form

\begin{equation}
A_\partial = \frac{{\cal A}(t) \,\mathrm{d} t}{\displaystyle \left(1+\frac{r^2}{\ell^2}\right)^{\frac{n}{2}}}
= \frac{a(t) \,\mathrm{d} t}{\displaystyle \left(1+r^2\right)^{\frac{n}{2}}}\,.
\label{eq:profile}
\end{equation}
where $\ell$ is a length scale and $n$ is an integer controlling the fall-off at large $r$. In the last step we have used the
  conformal invariance of the asymptotic boundary metric:  only the dimensionless product $a = {\cal A}\,\ell$ is physically meaningful so we can set $\ell=1$ without loss of generality.

 When the amplitude $a$ is constant, static zero temperature solutions were found in  \cite{Horowitz:2014gva} for various $n$. One family of such solutions describe static, self-gravitating electric fields in AdS.  This family extends from $a=0$, where it meets with pure Poincar\'e AdS, to a maximum amplitude $a=a_{\max}$, where a naked curvature singularity appears. $a_{\max}$ increases with $n$ but is always finite\footnote{Even if $A_\partial$ has  compact support, there is a maximum amplitude.}. In \cite{Horowitz:2016ezu}, it was shown that the singularity extends for all $a> a_{\max}$.

Now suppose $a(t)$ is initially zero, and increases to a constant value larger than $a_{\max}$. Since there is no smooth static endpoint, it is likely that the curvature will grow indefinitely.  In  \cite{Crisford:2017zpi}, the time dependent solution was found numerically for the case $n=1$ and it was shown that $F^2$ does indeed grow as a power of time. This produces increasing curvature not just near the axis of symmetry, but everywhere along the horizon. Interestingly enough, the intrinsic geometry of the horizon  does not become singular. It is derivatives off the horizon that became large. The solutions are axisymmetric, but nonaxisymmetric perturbations  do not affect the evolution. This is because they are clearly stable in pure AdS, so will decay away before $a(t)$ is turned on. Even though the curvature does not diverge in finite time, this clearly violates the spirit of cosmic censorship.

To see the effect of the weak gravity conjecture, we add a charged scalar field $\Phi$ with action
\begin{equation}
S_m = - \frac{1}{4\pi G}\int \mathrm{d}^4 x\,\sqrt{-g}\left[(\mathcal{D}_a \Phi)(\mathcal{D}^a \Phi)^\dagger +\,m^2\Phi \Phi^\dagger\right]\,,
\label{eq:action}
\end{equation}
where $\mathcal{D}_a = \nabla_a-i\,q\, A_a$, $m$ is the charged scalar field mass and $q$ its charge.
The weak gravity conjecture in AdS differs from the asymptotically flat version. To derive the new bound, we require that   extremal charged black holes can decay. In flat space, this is a purely quantum mechanical instability, but in AdS it turns out to give rise to a classical instability. This instability is the so called charged superradiant instability \cite{Starobinskil:1974nkd,Gibbons:1975kk} whose endpoint has been studied over the past decade \cite{Basu:2010uz,Bhattacharyya:2010yg,Dias:2011tj,Gentle:2011kv,Markeviciute:2016ivy,Bosch:2016vcp,Dias:2016pma}.

In AdS, the onset of this superradiant instability depends on the size of the black hole. To stay as close as possible to the original weak gravity conjecture, we consider an arbitrarily small black hole.  Superradiant scattering for a scalar field of mass $m$ and charge $q$ occurs if \cite{Starobinskil:1974nkd,Gibbons:1975kk}
\begin{equation}
0<{\omega}<q\,\mu\,,
\label{eq:super}
\end{equation}
where ${\omega}$ is the frequency of the perturbation we are considering and $\mu$ is the difference between $A_t$ at infinity and on the horizon.  To leading order in the size of the black hole, small extremal black holes have $\mu=1$ and the minimum possible frequency is given by the normal mode in AdS: ${\omega}= \Delta $ where
\begin{equation}\label{eq:Delta}
\Delta = \frac{3}{2}+\sqrt{\frac{9}{4}+m^2}\,.
\end{equation}
In the context of gauge/gravity duality, $\Delta$ is the conformal dimension of the operator dual to $\Phi$.
Substituting in Eq.~(\ref{eq:super}), gives the following lower bound on the scalar field charge $q$
\begin{equation}
q \ge q^{W}\equiv {\Delta}\,.
\end{equation}
This is the weak gravity bound in AdS. 

 It was shown in  \cite{Horowitz:2016ezu} that if $q \ge q^W$, the static solutions with constant $a$ become unstable to turning on $\Phi$ as $a$ is increased: for all profiles $n$, $\Phi$ perturbations become unstable before $a$ reaches $a_{\max}$. It was also shown that once $\Phi$ is nonzero, there is no maximum amplitude, so cosmic censorship cannot be violated as before. The previous solutions also become unstable for $q$ slightly less than $q^W$, but in this case, the solution with scalar field included becomes singular as $a$ increases, so one could again violate cosmic censorship. Thus, the minimum $q$ to preserve cosmic censorship is precisely the weak gravity bound. Note that we do not require that our field with $q>m$ has the smallest charge for this gauge field. If there was another field with $q\ll m$, it would simply remain zero and not affect the evolution.
 
 In \cite{Crisford:2018qkz} an  attempt was made to find analogous vacuum counterexamples to cosmic censorship in AdS. The boundary metric was chosen to take the form
 \begin{equation}
\mathrm{d} s^2_\partial = -\mathrm{d} t^2+\mathrm{d} r^2+r^2 [\mathrm{d}\phi - \tilde \omega(r) \mathrm{d} t]^2\;,
\label{eq:bndmetric}
\end{equation}
with
\begin{equation}
\tilde \omega(r) = a\,p(r).
\end{equation}
These metrics describe geometries with differential rotation with an amplitude $a$ and profile $p(r)$ that vanishes as $r\to \infty$. It was again found that smooth stationary solutions exist only up to a maximum amplitude $a_{\max}$. This is true both at zero and nonzero temperature.   To try to violate cosmic censorship, one can work at zero temperature and let the amplitude increase from zero to a value greater than $a_{\max}$.  Since there is no smooth stationary final state, it is likely that the curvature will grow without bound. The weak gravity bound cannot remove these examples since there is no Maxwell field.

However there is a problem with these vacuum examples. Before reaching $a_{\max}$, the boundary metric develops an ergoregion which extends into the bulk. The amplitude where the ergoregion first forms depends on the profile, but it is always less than $a_{\max}$. Since there are negative energy excitations in the ergoregion, the total energy can be reduced. It was argued in \cite{Crisford:2018qkz} that with  boundary metrics of this type, the energy is likely to be unbounded from below. Since one usually requires theories to have a minimum energy ground state, these 
examples are not on the same footing as the electromagnetic ones. 

Thus, at present, the strongest counterexamples to cosmic censorship involve a Maxwell field and are removed by assuming the existence of charged matter satisfying the weak gravity bound. Below we will present further evidence for a deep connection between the cosmic censorship and weak gravity conjectures.

\section{The Dilatonic Case}

\subsection{Charged dilatonic  black holes and the weak gravity bound }
Supergravity and string theory  contain dilatons, \ie, neutral scalar fields with exponential coupling to matter fields.
We will consider the following action
\begin{equation}
S=\frac{1}{16\pi G}\int \mathrm{d}^4x\sqrt{-g}\left[R+{6}-e^{-\,2\,\alpha\,\phi}F^{ab}F_{ab}-2\,\nabla_a \phi \nabla^a \phi\right]\,,
\label{eq:action1}
\end{equation}
where  $\phi$ is the dilaton and $\alpha\in\mathbb{R}$ is a dilatonic coupling.
The equations of motion derived from the action (\ref{eq:action1}) can be recast in the following form
\begin{subequations}\label{eqs:EOM}
\begin{align}
&R_{ab}+{3}g_{ab}=2\,\nabla_a \phi \nabla_b \phi+2\,e^{-\,2\,\alpha\,\phi}\left(F_{a}^{\phantom{a}c}F_{bc}-\frac{1}{4}g_{ab} F^{cd}F_{cd}\right)\,,\label{eq:einsteindilaton}
\\
& \nabla_a\left(e^{-\,2\,\alpha\,\phi}F^{ab}\right)=0\,,
\\
&\nabla_a\nabla^a \phi+\frac{\alpha}{2}e^{-\,2\,\alpha\,\phi}F^{ab}F_{ab}=0\,.\label{eq:dilaton}
\end{align}
\end{subequations}
The system of PDEs (\ref{eqs:EOM}) enjoys the following discrete symmetry $(\phi,\alpha)\leftrightarrow -(\phi,\alpha)$, that we can use to take $\alpha\geq0$ without loss of generality. Also, we note that when $\alpha=0$, we can consistently set $\phi=0$, in which case Eq.~(\ref{eq:action1}) reduces to the Einstein-Maxwell action which was studied in great detail in \cite{Horowitz:2014gva,Horowitz:2016ezu,Crisford:2017zpi,Crisford:2017gsb} and reviewed in the previous section.

The weak gravity bound for theories with dilatons was discussed in \cite{Heidenreich:2015nta}. To motivate their condition, consider the static asymptotically flat black hole solutions to the equations (\ref{eqs:EOM}) without the cosmological constant. These were found in \cite{Gibbons:1987ps, Garfinkle:1990qj}
and take the form
\be
ds^2 = -f(r) dt^2 + f(r)^{-1} dr^2 + R(r)^2 d\Omega
\ee 
where 
\be 
f(r) = \left( 1-\frac{r_+}{r}\right)\left( 1-\frac{r_-}{r}\right)^{\frac{1-\a^2}{1+\a^2}},  \qquad
R(r) = r \left(1-\frac{r_-}{r}\right)^{\frac{\a^2}{1+\a^2}}
\ee
The vector potential is $A = (Q/r)dt$ and the dilaton is
\be
e^{ \phi} =\left(1-\frac{r_-}{r}\right)^{\frac{\a}{1+\a^2}}
\ee
The mass and charge of these black holes are
\be
M = \frac{r_+}{2} + \left( \frac{1-\a^2}{1+\a^2}\right) \frac{r_-}{2}, \qquad Q=\left( \frac{r_+r_-}{1+\a^2}\right)^{1/2}
\ee  
and their Hawking temperature is
\be
T = \frac{1}{4\pi r_+}\left(\frac{r_+-r_-}{r_+}\right)^\frac{1-\a^2}{1+\a^2}
\ee

The extremal limit corresponds to $r_+ = r_-$, which implies
\be
\frac{Q^2}{M^2} = 1 + \a^2
\ee
Note that in the extremal limit, the horizon shrinks to zero size and the solution is  singular. The dilaton diverges there, but $F^2$ remains finite. For $\a < 1$, $T\to 0$ in the extremal limit as usual, but for $\a =1$ it remains constant and for $\a>1$ it diverges. Some implications of this unusual behavior are discussed in \cite{Holzhey:1991bx}.

The weak gravity bound should be the condition for (near) extremal black holes to exhibit charged superradiance. For a charged scalar field that is not directly coupled to the dilaton, this condition is still $\omega < q\mu$. But now $\mu = Q/r_+ = (1+\a^2)^{-1/2}$. So in asymptotically flat spacetime, the weak gravity bound would be $q/m >  (1+\a^2)^{1/2}$ since the minimum frequency is $\omega = m$. Note that this is just the statement that one needs a particle that exceeds the extremality bound for black holes. However, as explained earlier,  in AdS the minimum frequency is not $m$ but $\Delta$. So the bound is
\be\label{eq:dilbd}
q \ge q^W \equiv \Delta (1+\a^2)^{1/2}
\ee

\subsection{Results without charged matter\label{sec:resultsneutraldilatonic}}

\begin{figure}[h]
\centering
\includegraphics[width=0.45\linewidth]{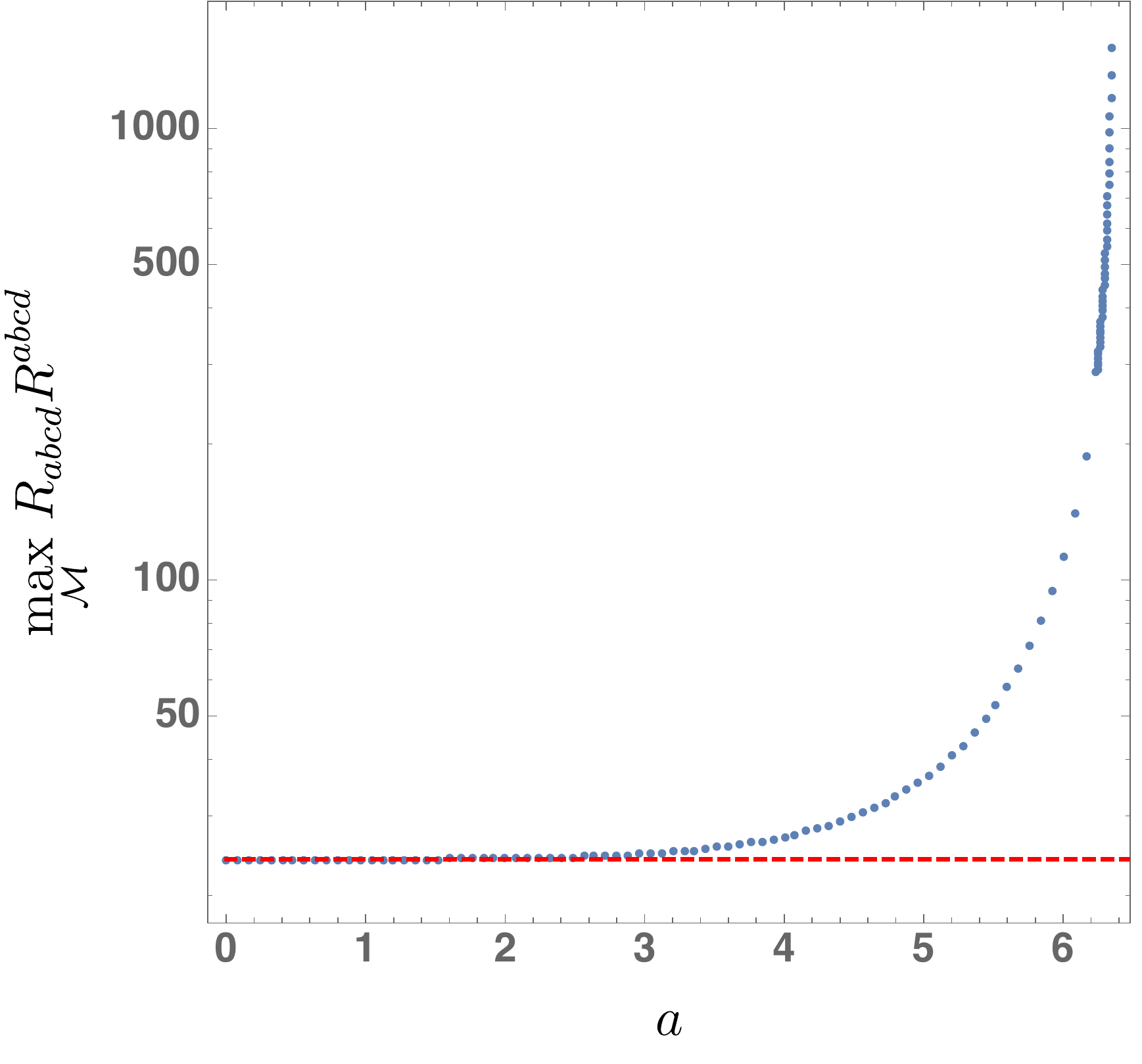}
\caption{Logarithmic plot of the maximum of the Kretschmann scalar, over the whole spacetime, as a function of the boundary amplitude $a$ for $\alpha=1$. Different values of $\alpha$ show a similar qualitative behaviour. The red dashed line denotes the curvature of pure AdS.} 
\label{fig:blow}
\end{figure}

We numerically constructed static, zero temperature solutions to this theory with a flat boundary metric and 
 boundary vector potential
\begin{equation}
A_\partial 
= \frac{a \,\mathrm{d} t}{\displaystyle \left(1+r^2\right)^{4}}\,.
\label{eq:profile4}
\end{equation}
(Other choices of fall-off yield similar results.) In the Appendix we briefly describe the numerical method used to obtain these solutions.
We again investigated whether a maximum amplitude exists when there is a dilatonic coupling. This turns out to be the case, as can be observed in Fig.~\ref{fig:blow}, where we plot in a logarithmic scale the maximum of the Kretschmann scalar, over the whole spacetime, as a function of the boundary amplitude $a$. For this case, we used $\alpha=1$, but similar results hold for different values of $\alpha$. The solution appears to develop a singularity as we approach $a_{\max}$.

\begin{figure}[h]
\centering
\includegraphics[width=0.45\linewidth]{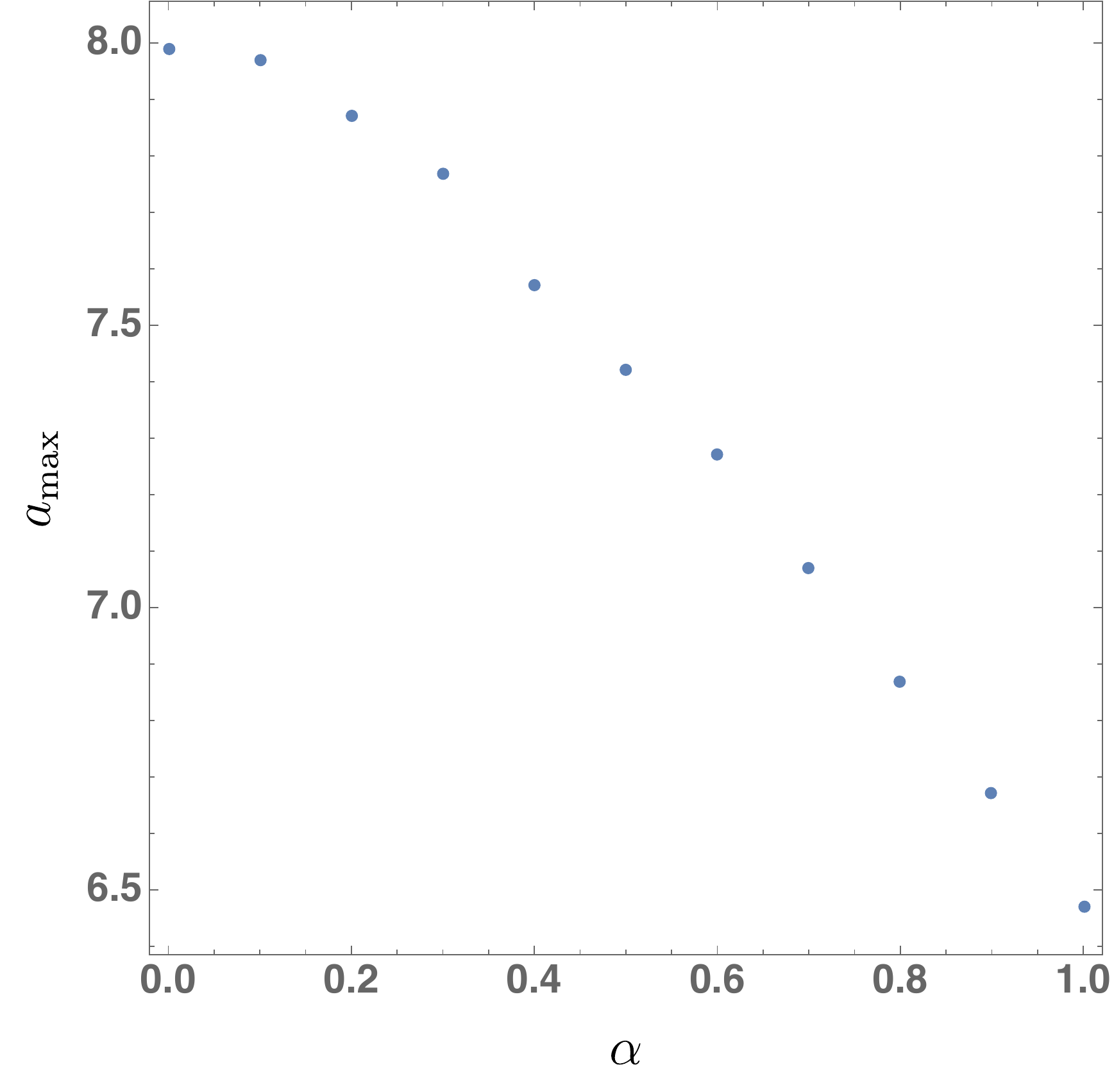}
\caption{$a_{\max}$ as a function of $\alpha$: increasing $\alpha$ decreases $a_{\max}$.}
\label{fig:amax}
\end{figure}

Next we studied how $a_{\max}$ depends on $\alpha$. It turns out that as $\alpha$ increases, we see that $a_{\max}$ decreases. This is exemplified in Fig.~\ref{fig:amax}, where we plot $a_{\max}$ as a function of $\alpha$. This is perhaps expected since the electromagnetic contribution to the stress energy tensor is enhanced by a $e^{-2\alpha \phi}$ factor with respect to the $\alpha=0$ case, \ie, pure Maxwell case. We note that for positive $\alpha$, $\phi$ must be negative, because of a simple maximum principle argument:  Since $A$ only has a time component, $F^{ab}F_{ab}\le 0$ and Eq. (\ref{eq:dilaton}) implies $\nabla^2 \phi \le 0 $. Since $\phi$ vanishes on all boundaries, it follows that it cannot have any local positive maximum, so it must be negative everywhere. 
Naturally, we have checked this to be true for all our solutions. However this argument is not the whole story, since we will soon see that the Maxwell field itself is smaller (for the same source) when the dilaton is present.

As before, if we now allow the amplitude to be time dependent and grow from $a=0$ to $a > a_{\max}$    we expect the curvature to grow without bound. Thus  these examples provide a new  class of theories where one can violate weak cosmic censorship. Next, we would like to understand whether the inclusion of a charged scalar field can save cosmic censorship.

\subsection{Results with charged matter}
We now add to Eq.~(\ref{eq:action1}) a charged scalar field $\Phi$ with charge $q$ and mass $m$:
\begin{equation}
S=\frac{1}{16\pi G}\int \mathrm{d}^4x\sqrt{-g}\left[R+{6}-e^{-\,2\,\alpha\,\phi}F^{ab}F_{ab}-2\,\nabla_a \phi \nabla^a \phi-4(\mathcal{D}_a\Phi)^\dagger(\mathcal{D}^a\Phi)-4\,m^2\,\Phi^\dagger \Phi\right]\,,
\label{eq:action2}
\end{equation}
where $\mathcal{D}=\nabla -i\,q\,A$ and $q$ is the scalar field electric charge. The  new equations of motion read
\begin{subequations}
\label{eqs:EOMC}
\begin{align}
&R_{ab}+{3}g_{ab}=2\,\nabla_a \phi \nabla_b \phi+2\,e^{-\,2\,\alpha\,\phi}\left(F_{a}^{\phantom{a}c}F_{bc}-\frac{1}{4}g_{ab} F^{cd}F_{cd}\right)
\\
&\hspace{4cm}+2(\mathcal{D}_a\Phi)(\mathcal{D}_b\Phi)^\dagger+2(\mathcal{D}_a\Phi)^\dagger(\mathcal{D}_b\Phi)+2\,m^2\,g_{ab}\Phi^\dagger \Phi\,,\nonumber
\\
& \nabla_a\left(e^{-\,2\,\alpha\,\phi}F^{a}_{\phantom{a}b}\right)=i\,q\,[(\mathcal{D}_b\Phi)\Phi^\dagger-(\mathcal{D}_b\Phi)^\dagger\Phi]\,,
\\
&\nabla_a\nabla^a \phi+\frac{\alpha}{2}e^{-\,2\,\alpha\,\phi}F^{ab}F_{ab}=0\,,
\\
& \mathcal{D}_a\mathcal{D}^a\Phi = m^2\Phi\,.\label{eq:chargedscalar}
\end{align}
\end{subequations}

To check whether the  solutions discussed in section \ref{sec:resultsneutraldilatonic} are stable to turning on $\Phi$, we will first consider the case where $\Phi$ is perturbatively small, such that its backreaction on the metric, gauge field and dilaton is negligible. In this case, we merely search for linear solutions of Eq.~(\ref{eq:chargedscalar}) around the above backgrounds, for several values of $m$. Instead of $m$, we will parametrise our solutions by $\Delta$ (\ref{eq:Delta}).
 Finiteness of energy requires $\Delta\geq1$, with saturation occurring at the so called unitarity bound. We considered both $\Delta = 4$ and $\Delta =2$ corresponding to $m^2 = 2$ and $m^2 = -2$, and obtained similar results in all our studies.

We are interested in studying the onset of scalar condensation around our backgrounds as a function of the boundary amplitude $a$. These perturbative solutions are independent of time and axisymmetric. We thus view Eq.~(\ref{eq:chargedscalar}) as an eigenvalue equation for  $q^2$, for given value of $a$.

\begin{figure}[h]
\centering
\includegraphics[width=0.55\linewidth]{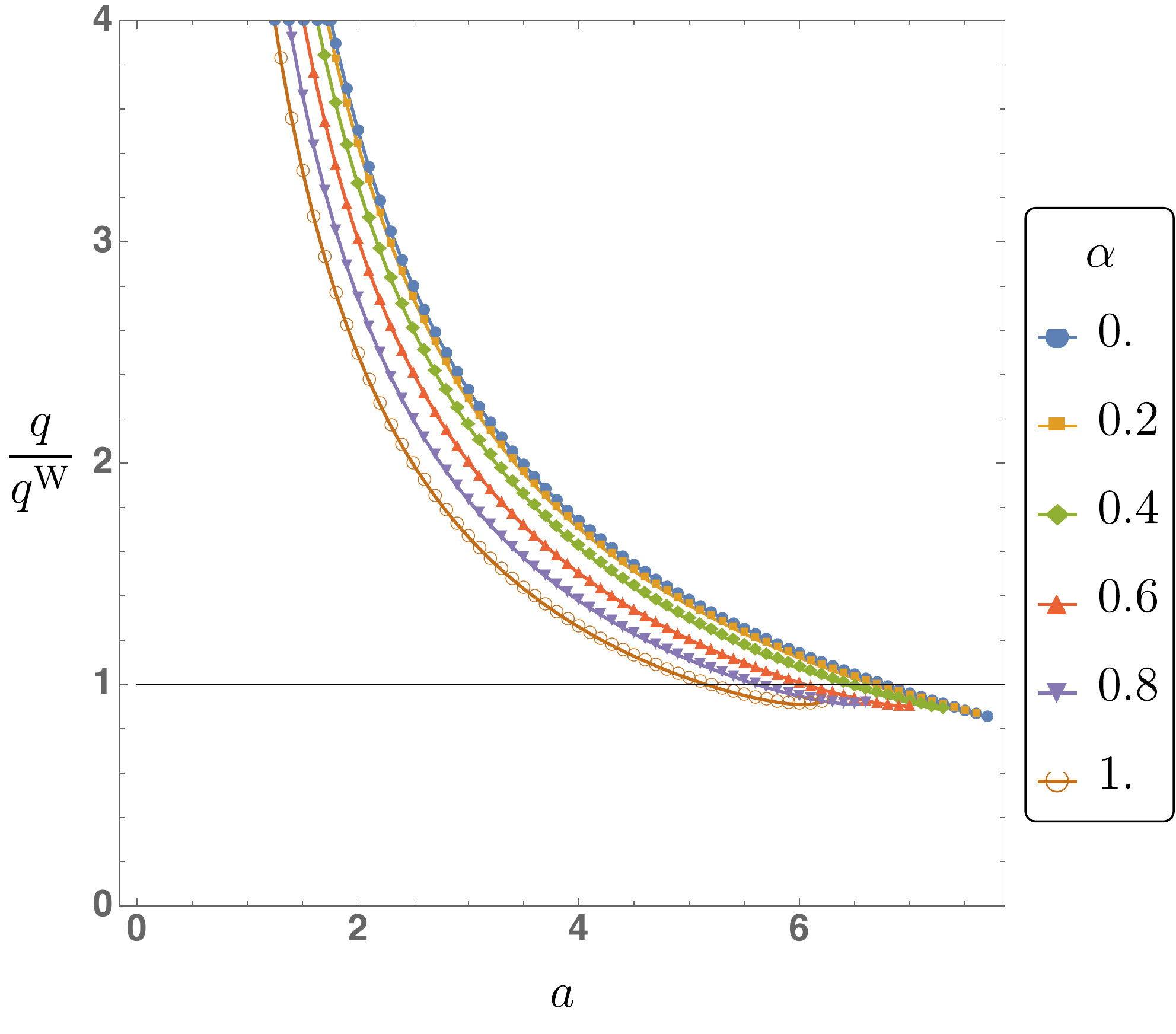}
\caption{Minimal charge $q$ for condensing $\Phi$ as a function of the boundary amplitude $a$ for several values of the dilaton coupling $\alpha$, which are labelled on the right.}
\label{fig:lineardilatonic}
\end{figure}

First, we studied the linear results for $\Delta = 4$. We find that for all $\alpha$, there exists a critical value of $q$ above which the dilatonic solutions become unstable to charged scalar perturbations. For a fixed amplitude on the boundary, the critical value of $q$ increases with $\alpha$, since the electric field is reduced. 
 Furthermore, as we approach $a_{\max}$, this critical value always drops below the weak gravity bound. 
 This is shown in Fig.~\ref{fig:lineardilatonic}, where we plot $q/q^{\mathrm{W}}$ as a function of $a$ for several values of $\alpha$, which are labelled on the figure. This means that if the weak gravity bound is satisfied, our dilatonic counterexamples to cosmic censorship are not valid, and we must study the solutions with the charged scalar included.
This is not true if we use the pure Maxwell bound $q\geq \Delta$.  For $\alpha\approx 1$, the dilatonic solutions remain stable for some charges satisfying this condition. One needs the stronger bound $q > q^{\mathrm{W}}$ (\ref{eq:dilbd}) predicted for dilatonic theories to ensure instability.
 Similar results hold for any value of $\Delta$ we have tried.

\begin{figure}[h]
\centering
\includegraphics[width=0.45\linewidth]{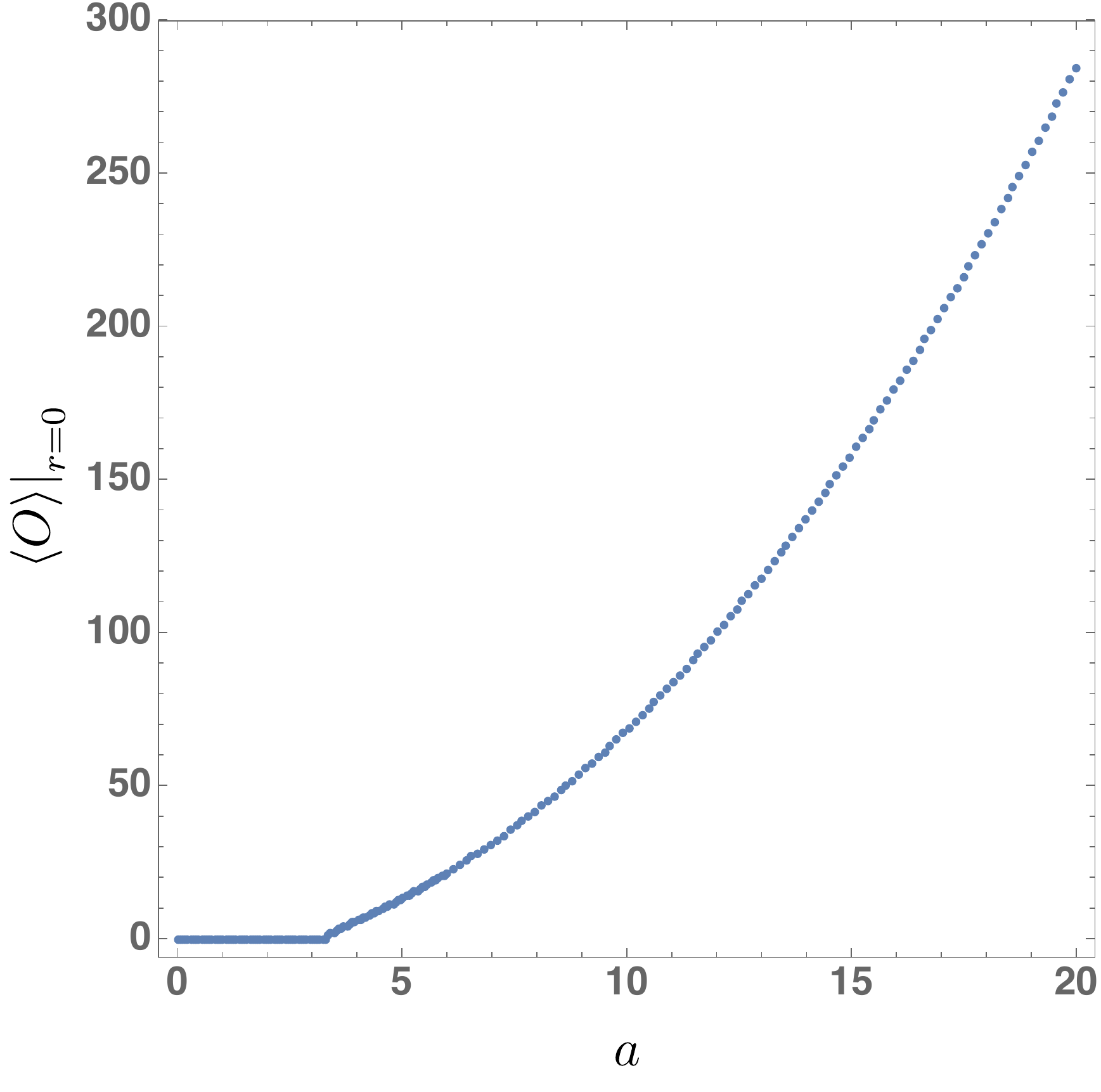}
\caption{Central value of expectation value of the operator dual to $\Phi$ as a function of the boundary amplitude $a$ for $\alpha = \sqrt{3}$. The condensation occurs around $a\sim 3.33$, in accordance with the linear results. There is no maximum amplitude.}
\label{fig:largecond}
\end{figure}

Next, we constructed the full nonlinear solutions with the charged scalar included (see Appendix). For this we found it easier to use $\Delta = 2$, which we assume henceforth. If $q\ge q^{\mathrm{W}}$, we find no evidence of a maximum amplitude.
This is best illustrated by looking at the expectation value of the operator $\mathcal{O}$ dual to $\Phi$ evaluated at the origin on the boundary. In Fig.~\ref{fig:largecond} we plot this expectation value for $q=q^{\mathrm{W}}$ as a function of the amplitude $a$. We see the condensation starting around $a\sim 3.33$, and the solutions extending to values as large as $a\sim 20$. This figure was constructed for $\alpha = \sqrt{3}$, but similar behaviour occurs for different values of $\alpha$. Thus, if the weak gravity bound is satisfied, one cannot violate cosmic censorship this way.

\begin{figure}[h]
\centering
\includegraphics[width=0.45\linewidth]{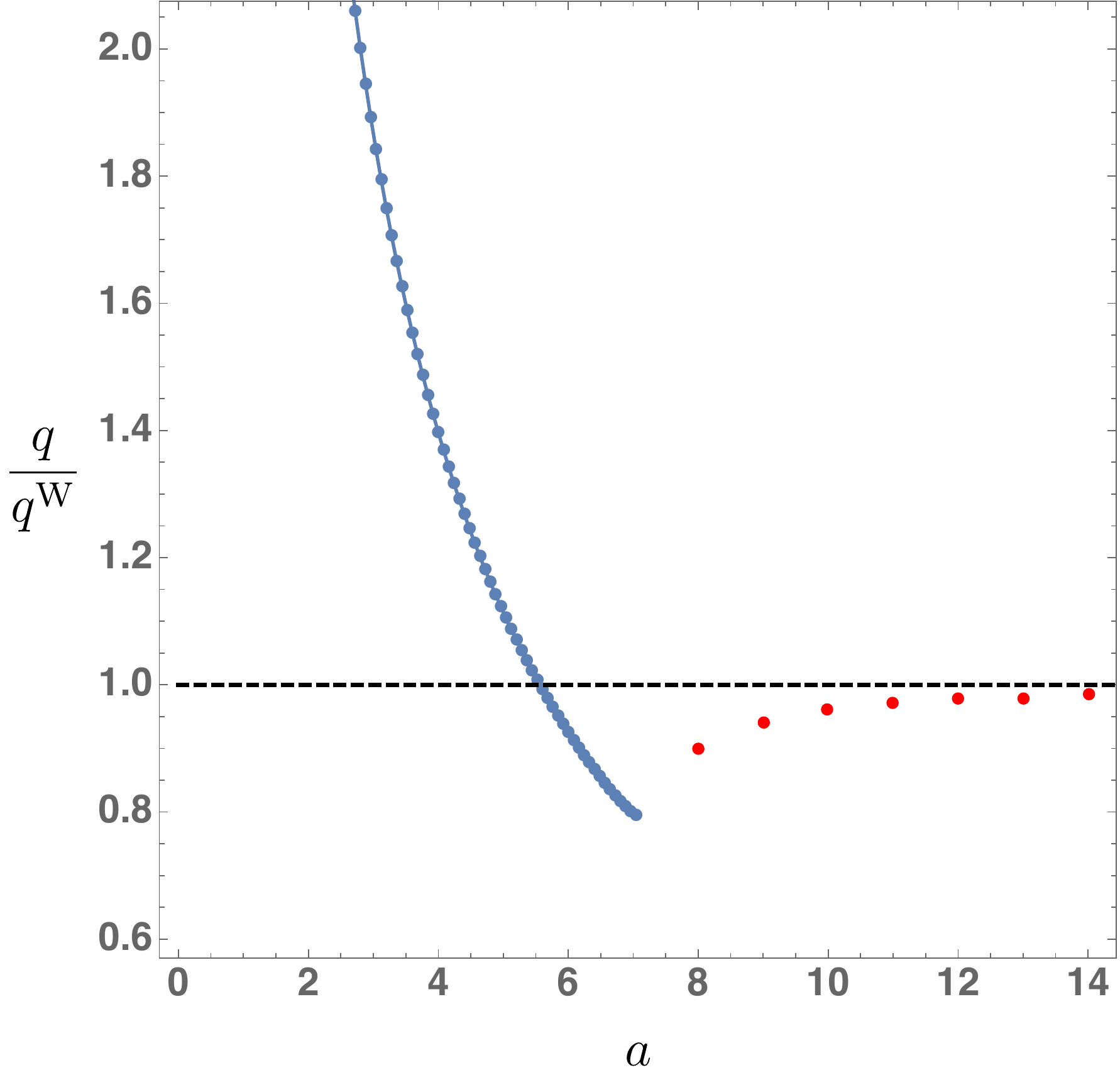}\hspace{0.5cm}\includegraphics[width=0.45\linewidth]{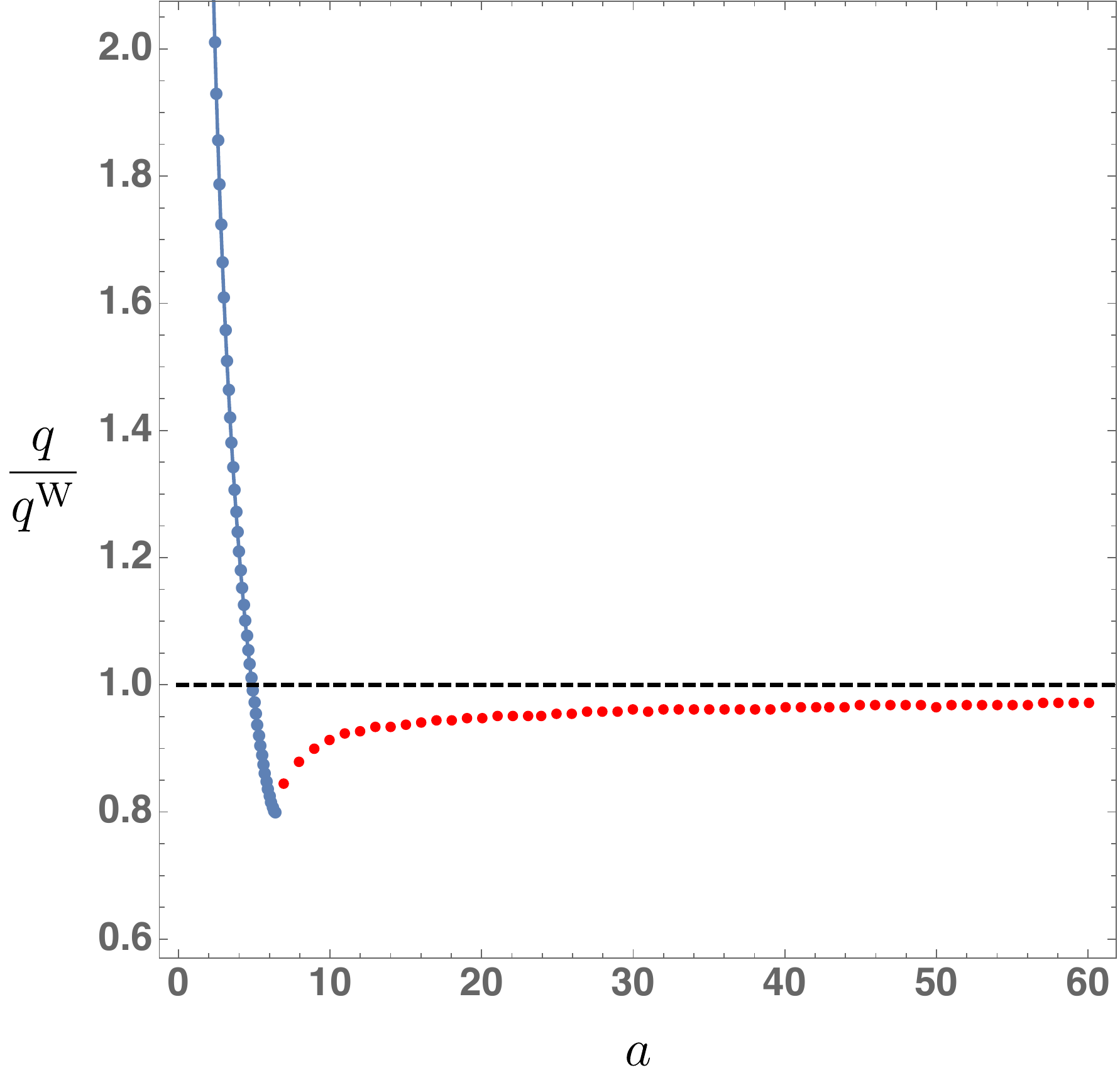}
\caption{For $\a <1$, the bound on $q/\Delta$ needed to preserve cosmic censorship is precisely the weak gravity bound: the blue solid line indicates the onset of solutions with $\Phi\neq0$, and the red dots show the approximate location of singular solutions. On the left plot we have $\alpha=1/\sqrt{3}$ and on the right $\alpha = 0.9$.  Notice the different scale on the horizontal axes. In both cases, $\Delta = 2$.}
\label{fig:crazy}
\end{figure}

Fig.~\ref{fig:lineardilatonic} shows that we can lower the charge slightly below the weak gravity bound and still turn on $\Phi$. However, if $\a < 1$, there is again a maximum amplitude where the solution becomes singular. To show this, we started with the $q= q^{\mathrm{W}}$ solutions with $a > a_{\max}$ and lowered $q$ until we found an obstruction.
 This is  illustrated in Fig.~\ref{fig:crazy} where we show the onset curve for $\Phi$ for $a<a_{\max}$, and the corresponding singular curve for $a>a_{\max}$. The two plots correspond to two different choices of dilaton coupling: $\a = 1/\sqrt 3$ and $\a = .9\ .$  (The value $\a = 1/\sqrt 3$ is obtained by dimensionally reducing a  five dimensional charged black string \cite{Gibbons:1994vm}.) Notice that both singular curves approach the weak gravity bound, but as $\a \to 1$, the approach becomes very slow. Whenever there is a maximum amplitude for smooth solutions, one can violate cosmic censorship by increasing the amplitude on the boundary past this bound. This shows that for $\alpha <1$, the weak gravity bound is precisely what is needed to preserve cosmic censorship.
 
\begin{figure}[h]
\centering
\includegraphics[width=0.45\linewidth]{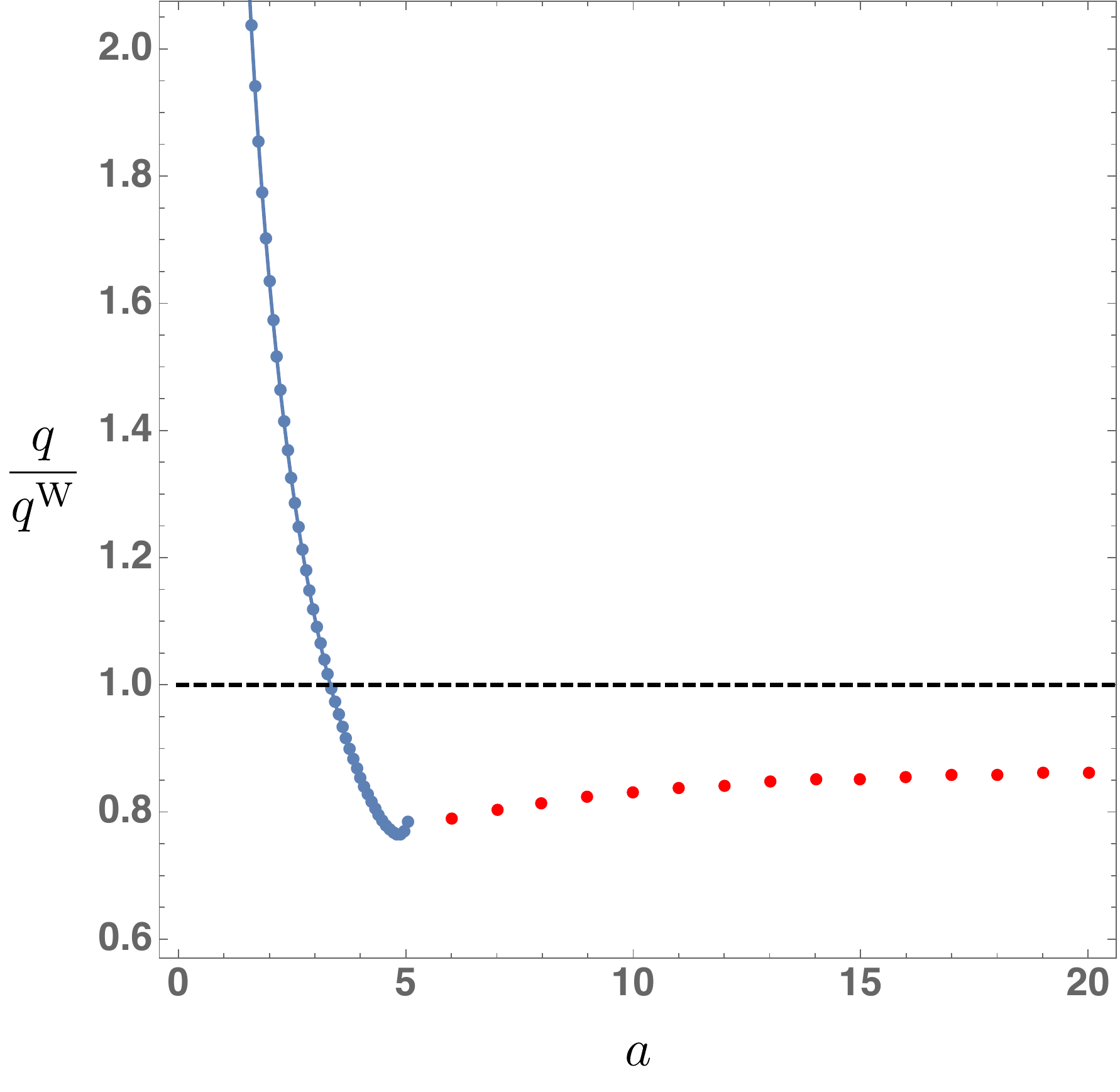}
\caption{The blue solid line indicates the onset of solutions with $\Phi\neq0$, and the red dots the approximate location of singular solutions. This plot was generated with $\alpha=\sqrt{3}$. Similar results hold for $\alpha\geq1$.}
\label{fig:crazy2}
\end{figure}

For $\alpha > 1$, the singular curve does not seem to approach the weak gravity bound. This is shown  in Fig.~\ref{fig:crazy2} which was computed for $\alpha = \sqrt{3}$. (This value is obtained by standard Kaluza-Klein reduction of a five dimensional vacuum solution.) It is possible that the singular curve will eventually approach $q / q^W =1$  at much larger amplitudes, but if not, cosmic censorship is preserved in these theories for charges slightly less than the weak gravity bound. The borderline case of $\a = 1$ is the value of the dilaton in string theory.  The plot in this case looks similar to the case $\a = 0.9$.

We do not understand the qualitative difference between $\a <1$ and $\a>1$. However it is worth recalling another qualitative difference between these two cases:  the temperature of black holes vanishes in the extremal limit if $\a <1$ and diverges if $\a>1$. It is tempting to think there might be a connection between these two facts, even though there are no black holes in our examples. It was shown in \cite{Holzhey:1991bx} that even though the temperature diverges when $\a > 1$, the flux of energy at infinity still goes to zero in the extremal limit. The black hole effectively becomes thermally insulated from the asymptotic region by large potential barriers. This means that extremal black holes will still not evaporate.

\begin{figure}[h]
\centering
\includegraphics[width=0.45\linewidth]{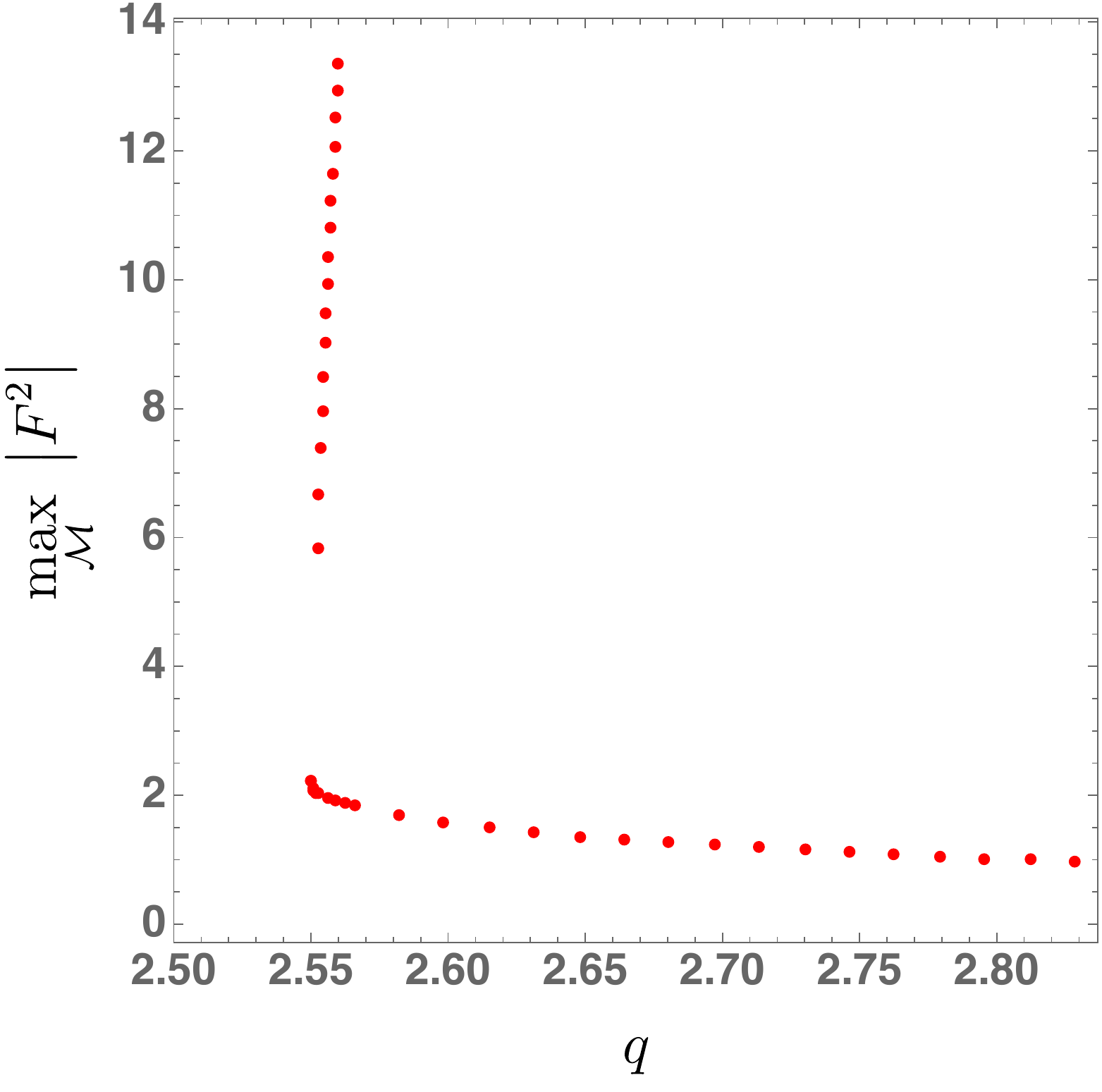}
\caption{Approach to the singular solution for fixed $\alpha=1$, and $a=10$: for a small range of $q$, there is a two-fold non-uniqueness.}
\label{fig:need}
\end{figure}

Finally, we note that the approach to the singularity is not monotonic in $q$. This is a new feature which does not seem to occur for $\alpha=0$. In fact, we suspect that the approach will exhibit a spiralling behaviour similar to the one observed in \cite{Markeviciute:2017jcp}. One quantity where this behaviour is apparent is the maximum value of $|F^2|$ over the whole spacetime, which we plot in Fig.~\ref{fig:need} for fixed $a=10$ and $\alpha=1$. This maximum value is finite at $q_{\min}$, but then grows rapidly on a second branch of solutions at slightly larger $q$.

\section{The Multi-Charged Case}
\subsection{The field equations and weak gravity bound}
In this section, we study the effect of two Maxwell fields  and their respective charged scalars,  without a dilaton. The new action reads
\begin{equation}
S=\frac{1}{16\pi G}\int \mathrm{d}^4 x\sqrt{-g}\left[R+{6}-\sum_{I=1}^{2} \left (F_{I}^{\;ab}F_{I\;ab}+4\,(\mathcal{D}_{I}^{\;a} \Phi_I)^\dagger(\mathcal{D}_{I\;a}\Phi_I)+4m^2_I \Phi_I^\dagger \Phi_I \right)\right]\,,
\end{equation}
where upper case latin indices label the fields, $F_I = \mathrm{d}A_I$, $\mathcal{D}_I = \nabla-i q_I\,\,A_I$ and $m^2_I=\Delta_I(\Delta_I-3)$, all for $I = 1,2$. Unless otherwise stated, repeated upper case latin indices should not be summed over.

The equations of motion are:
\begin{subequations}
\label{eqs:EOMD}
\begin{align}
&R_{ab}+{3}g_{ab}=2 \sum_{I=1}^2\left[F_{I\;a}^{\phantom{\;a}c}F_{I\;bc}-\frac{1}{4}g_{ab} F_{I}^{\;cd}F_{I\;cd}\right]\nonumber
\\
&\hspace{4cm}+2\sum_{I=1}^2\left[(\mathcal{D}_{I\;a}\Phi_I)(\mathcal{D}_{I\;b}\Phi_I)^\dagger+(\mathcal{D}_{I\;a}\Phi_I)^\dagger(\mathcal{D}_{I\;b}\Phi_I)+\,m_I^2\,g_{ab}\Phi_I^\dagger \Phi_I\right]\,,
\\
& \nabla_aF^{\;a}_{I\phantom{a}b}=i\,q_I\,[(\mathcal{D}_{I\;b}\Phi_I)\Phi_I^\dagger-(\mathcal{D}_{I\;b}\Phi_I)^\dagger\Phi_I]\,,
\\
& \mathcal{D}_{I\;a}\mathcal{D}_{I}^{\;a}\Phi_I = m^2_I\Phi_I\,.\label{eq:chargedscalarD}
\end{align}
\end{subequations}
For concreteness we will also use $\Delta_1=\Delta_2=2$, corresponding to the  masses $m_1^2=m_2^2=-2$.

The weak gravity bound must be modified when there is more than one type of charge. This can be seen as follows \cite{Cheung:2014vva}.
An extremal black hole with charges $Q_I$ can decay into particles with charges $q_I$ and masses $m_I$ only if $Q_I = n_I q_I$ and 
\be
\sum_I (n_I q_I)^2 = M^2 \ge \left(\sum_I  n_I m_I \right)^2
\ee
In our case when there are two types of charge, this implies that the charge to mass ratios $z_I = q_I/m_I$ satisfy 
\be\label{eq:twofieldbd}
z_1^2 + z_2^2 \le (z_1 z_2)^2\quad {\rm or\ equivalently}\quad \frac{1}{z_1^2} + \frac{1}{z_2^2} \le 1
\ee
Geometrically, this is the statement that the convex hull of the charge to mass ratios includes the unit disk \cite{Cheung:2014vva}. In AdS, the bound is identical with the understanding that $z_I = q_I/ \Delta_I$.

For simplicity, we take two scalar fields, but we suspect that our conclusions will remain unchanged for more charged scalar fields. We will also take the profile of the gauge fields at the boundary to have the same functional dependence in $r$, but different amplitudes. That is to say, we choose
\begin{equation}
A_{I\;\partial}=\frac{a_I}{(1+r^2)^4}\mathrm{d}t\,.
\label{eq:proT}
\end{equation}
We have tried a couple of different fall offs, and the results remain unchanged.

\subsection{Results without charged fields and their stability}

When $\Phi_I=0$, the solutions for any $a_1$ and $a_2$ can be determined from the results for the single field case  \cite{Horowitz:2014gva,Horowitz:2016ezu,Crisford:2017gsb} by symmetry. First we note that without any scalars, our action has a $U(1)$ symmetry which rotates the two Maxwell fields. This means that for any $a_1$, $a_2$, we can choose an angle $\theta$ so that  $a_1 \sin \theta = a_2 \cos \theta$. Then if we define
\begin{equation}
\hat A = A_1 \cos \theta  + A_2 \sin \theta 
\end{equation}
the orthogonal linear combination will have no source on the boundary and hence vanish everywhere.
Thus, our solutions reduce to the one charge case for $\hat A$ with amplitude $a = (a_1^2 + a_2^2)^{1/2}$. It follows that everywhere in the solution
\begin{equation}
A_1 = \hat A \cos \theta\,,\quad A_2 = \hat A \sin \theta\,.
\end{equation}
Clearly, smooth static solutions again only exist up to a maximum amplitude, and the counterexamples to cosmic censorship that we had before trivially extend to this theory (with $\Phi_I=0$).

When we add the two scalars with different charges we break the $U(1)$ symmetry. But to compute when these solutions become unstable,  we  treat the scalars as linear perturbations and can use the background symmetry. At linear order, \ie,   solving for Eq.~(\ref{eq:chargedscalarD}) around solutions with $\Phi_I=0$,  we have an equation depending on, \emph{e.g.}, $q_1 A_1 = q_1 \hat A \cos \theta$. For any amplitude for $\hat A$, the onset of the instability must satisfy
\begin{equation}
q_1 \cos \theta = q_2 \sin \theta \Rightarrow q_1 a_1 = q_2 a_2.
\end{equation}

This relation makes it easy to read off the onset of the instability of the two Maxwell field solutions from the known results for a single field. If the single field solution with amplitude $a_0$ becomes unstable for $q>q_0$, then the two charge solution with 
$a_1 = a_0 \cos\theta,\ a_2 = a_0\sin\theta$ becomes unstable for $q_1 > q_0 /\cos\theta$ or $ q_2 > q_0/\sin\theta$. Thus the onset of the instability for solutions with $a_1^2+a_2^2 = a_0^2$ satisfies
\be\label{eq:singular}
 \left (\frac{q_0}{q_1}\right)^2 + \left (\frac{q_0}{q_2}\right)^2 = 1
 \ee
In particular, this is true at the maximum amplitude $a_{\max}$.
For a profile such as (\ref{eq:proT}) for the single field case,  we find $q_{0}\approx 1.38564$ for $a_{\max}\approx 7.97$. \cite{Horowitz:2014gva,Crisford:2017gsb}.

\begin{figure}[h]
\centering
\includegraphics[width=0.45\linewidth]{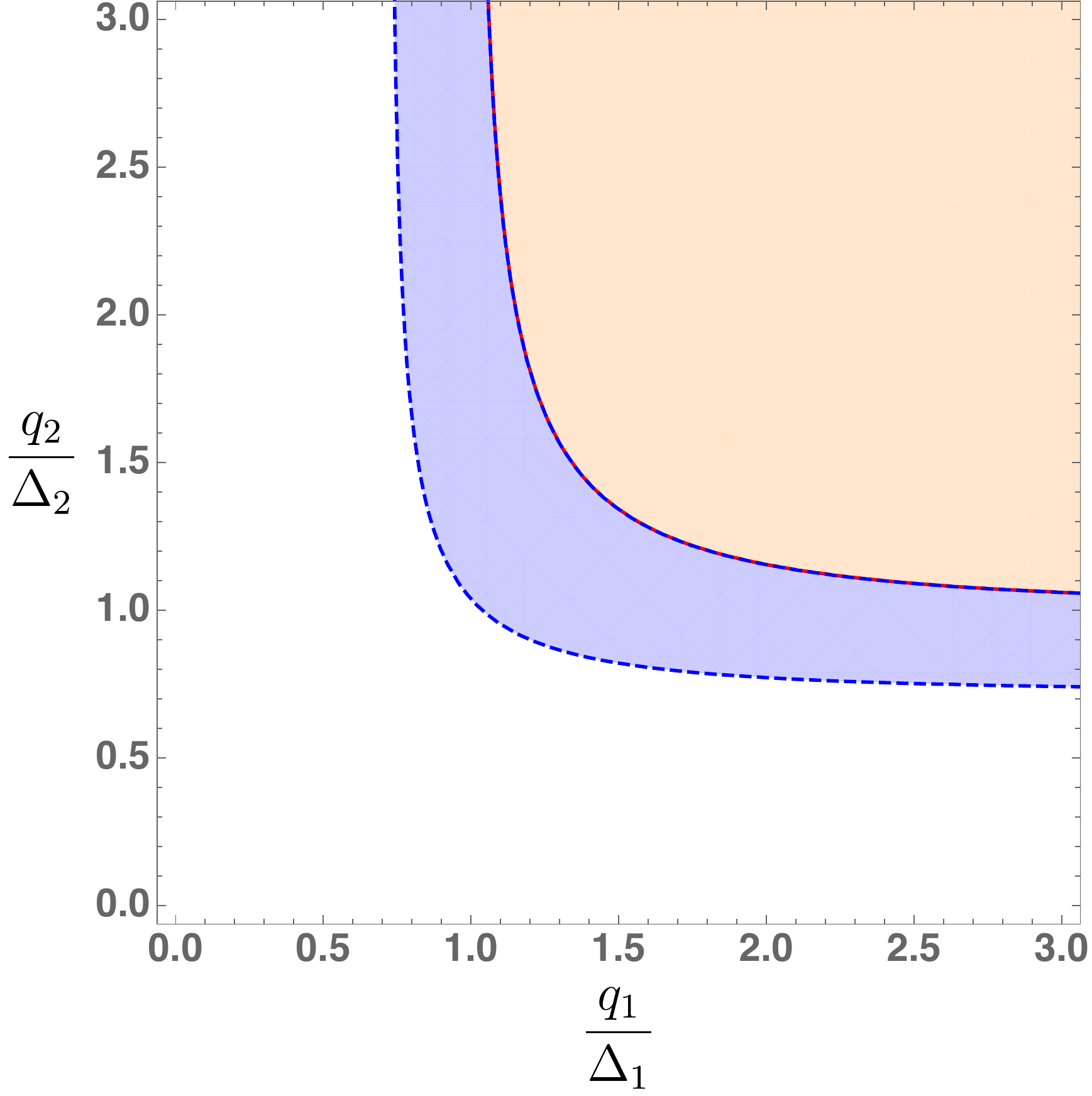}
\caption{The weak gravity bound  (\ref{eq:twofieldbd}) is satisfied in the orange region. The boundary of the blue region denotes the stability boundary (for $\Delta_1 = \Delta_2 = 2$) in the sense that the solutions remain stable for all  $a< a_{\max}$ only for charges below (or to the left of) this boundary. Since these regions do not intersect, our solutions always  become unstable before reaching $a_{\max}$.}
\label{fig:analytic}
\end{figure}

  This result of linear stability is illustrated in Fig.~\ref{fig:analytic}. The top orange region is the region satisfying the weak gravity bound (\ref{eq:twofieldbd}). The lower boundary of the blue region is the curve  (\ref{eq:singular}) with $q_0 \approx 1.38564$ corresponding to $a = a_{\max}$.  For smaller amplitudes, this threshold curve moves up into the blue and orange regions. Note that if a point $(\tilde q_1, \tilde q_2)$ is the onset of the instability for a particular solution with amplitudes $a_I$, that solution is stable only if both charges are less than these values, \ie, only if the charges lie in the bottom rectangle $0\le q_1\le \tilde q_1$ and $  0\le q_2\le \tilde q_2$. Once we reach $a_{\max}$, this bottom rectangle never intersects the orange region showing that if the weak gravity bound is satisfied, the solutions always become unstable before reaching $a_{\max}$.
 
\subsection{Results with charged matter}

We again numerically solved the full nonlinear equations with scalars, looking for zero temperature static solutions.
We took data of the form $a_1=\lambda\,a_2$ for different values of $\lambda>1$\footnote{Values of $\lambda<1$ can be recovered by exchanging $(\Phi_1,A_1)\leftrightarrow(\Phi_2,A_2)$.}. By virtue of this choice, $a_1$ will be larger than $a_2$, so it will always be easier to condense $\Phi_1$ than $\Phi_2$, \ie, the minimum value of $q_1$ necessary to condense $\Phi_1$ will always be smaller than the value of $q_2$ needed to condense $\Phi_2$.  
We first explore whether there is still a maximum amplitude. So
we fix $\lambda >1$ and take a pair $(q_1,q_2)$ saturating the the weak gravity bound. We then increase the amplitude $a_2$ keeping $a_1=\lambda\,a_2$. 
 There are three different phases: an initial phase where both $\Phi_I = 0$, a second phase where one scalar turns on but the other is still zero, and a final phase with both $\Phi_I\ne 0$. 
 
\begin{figure}[h]
\centering
\includegraphics[width=0.45\linewidth]{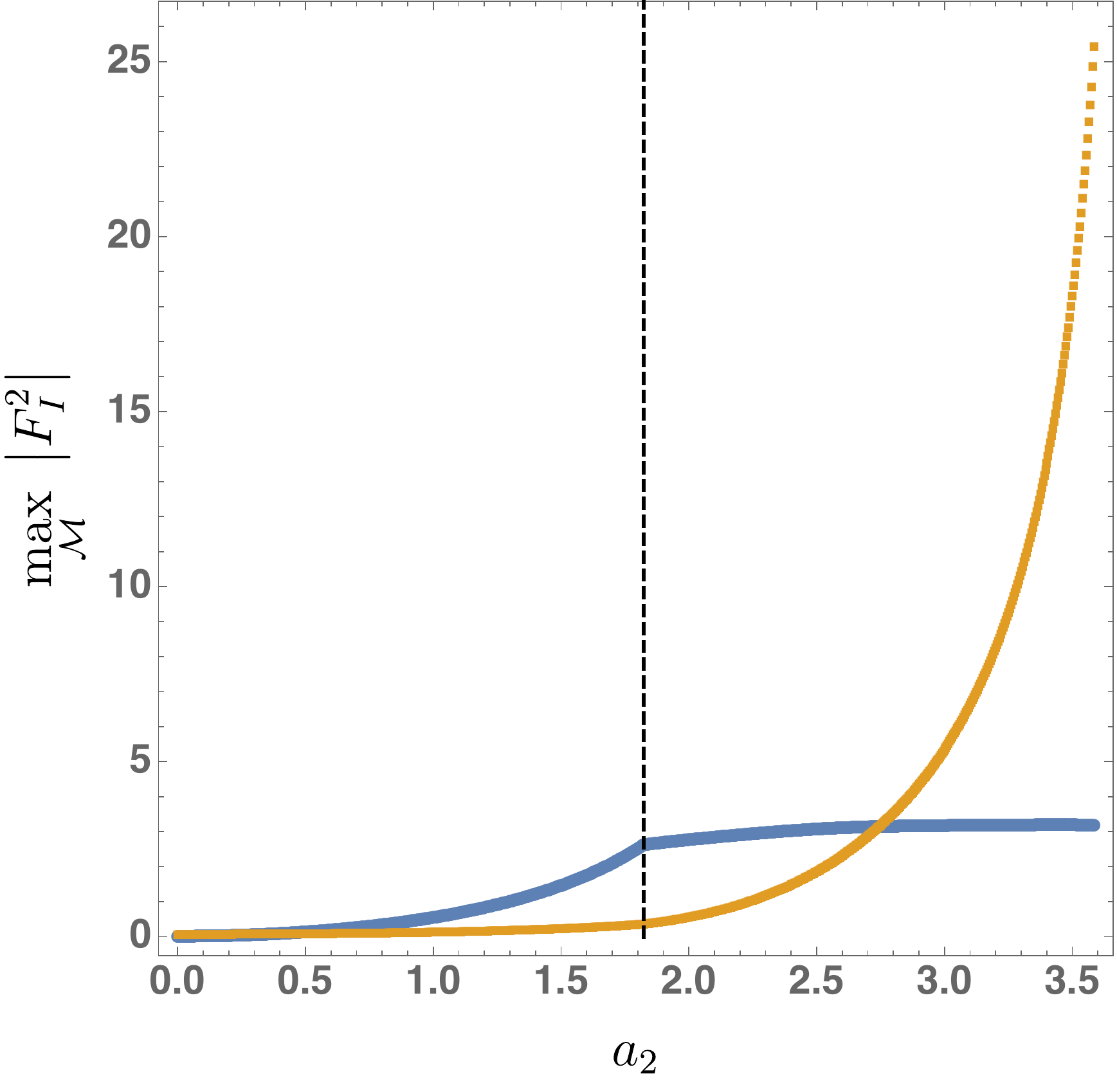}
\caption{$|F^2_I|$ as a function of $a_2$, for $a_1=3\,a_2$, $q_1=4/\sqrt{3}$ and $q_2=4$. The blue disks correspond to $|F_1^2|$, and the orange squares to $|F_2^2|$. The vertical dashed black line marks the onset of $\Phi_1$.}
\label{fig:nosecreening}
\end{figure}

If we allow only the first scalar to condense, the solution eventually becomes singular at some critical value of $a_2=a_2^\star>a_{2\;\max}$: the reason being that there is no charged scalar field to shield the electric field created by the second Maxwell field. This is illustrated in Fig.~\ref{fig:nosecreening} where we fixed $q_1=4/\sqrt{3}$, $q_2=4$ and $\lambda=3$. With these values, $\Phi_1$ condenses first.  The plot shows the maximum value of both $|F_1^2|$ and $|F_2^2|$ as we increase $a_2$. The vertical line marks the onset of condensation of $\Phi_1$, before which the solution has vanishing scalar fields, and after which $\Phi_1\neq0$ but $\Phi_2=0$. The blue disks represent $|F_1^2|$ and the orange squares $|F_2^2|$. As $a_2$ increases, $\Phi_1$ condenses and screens the electric field due to $A_1$, but there is nothing to screen $A_2$, since we have enforced  $\Phi_2=0$. Eventually, $|F_2^2|$ becomes too large, leading to an appearance of a singularity in the bulk, which is signalled by the fact that $|F_2^2|$ appears to diverge. Notice that after $\Phi_1$ condenses, $|F_1^2|$ stops growing and remains approximately constant.

\begin{figure}[h]
\centering
\includegraphics[width=0.45\linewidth]{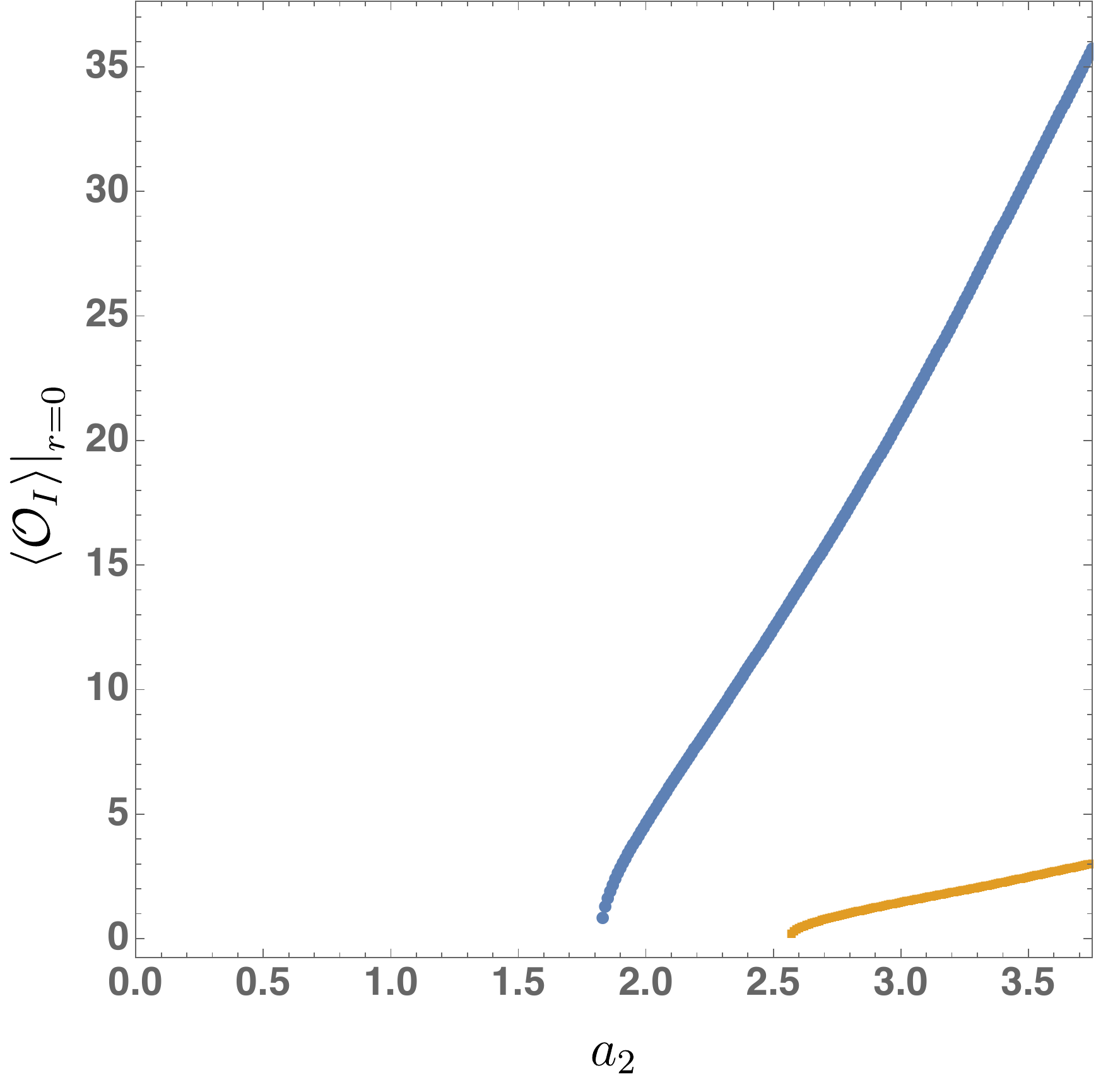}
\caption{$\left.\langle \mathcal{O}_I\rangle\right|_{r=0}$ as a function of $a_2$, for $a_1=3\,a_2$, $q_1=4/\sqrt{3}$ and $q_2=4$. The blue disks correspond to $\left.\langle \mathcal{O}_1\rangle\right|_{r=0}$, and the orange squares to $\left.\langle \mathcal{O}_2\rangle\right|_{r=0}$.}
\label{fig:twocon}
\end{figure}

However, if we do not insist that $\Phi_2 = 0$, before we reach $a_2^\star$, $\Phi_2$ will also condense, and this phase with \emph{two} condensed scalars  appears to extend to arbitrarily large  amplitude $a_2$. This can be seen in Fig.~\ref{fig:twocon} where we plot the expectation values of the operators dual to $\Phi_I$: $\left.\langle \mathcal{O}_1\rangle\right|_{r=0}$ and $\left.\langle \mathcal{O}_2\rangle\right|_{r=0}$ as a function of $a_2$ for $q_1=4/\sqrt{3}$, $q_2=4$ and $\lambda=3$. There is no sign of a maximum amplitude.

\begin{figure}[h]
\centering
\includegraphics[width=0.45\linewidth]{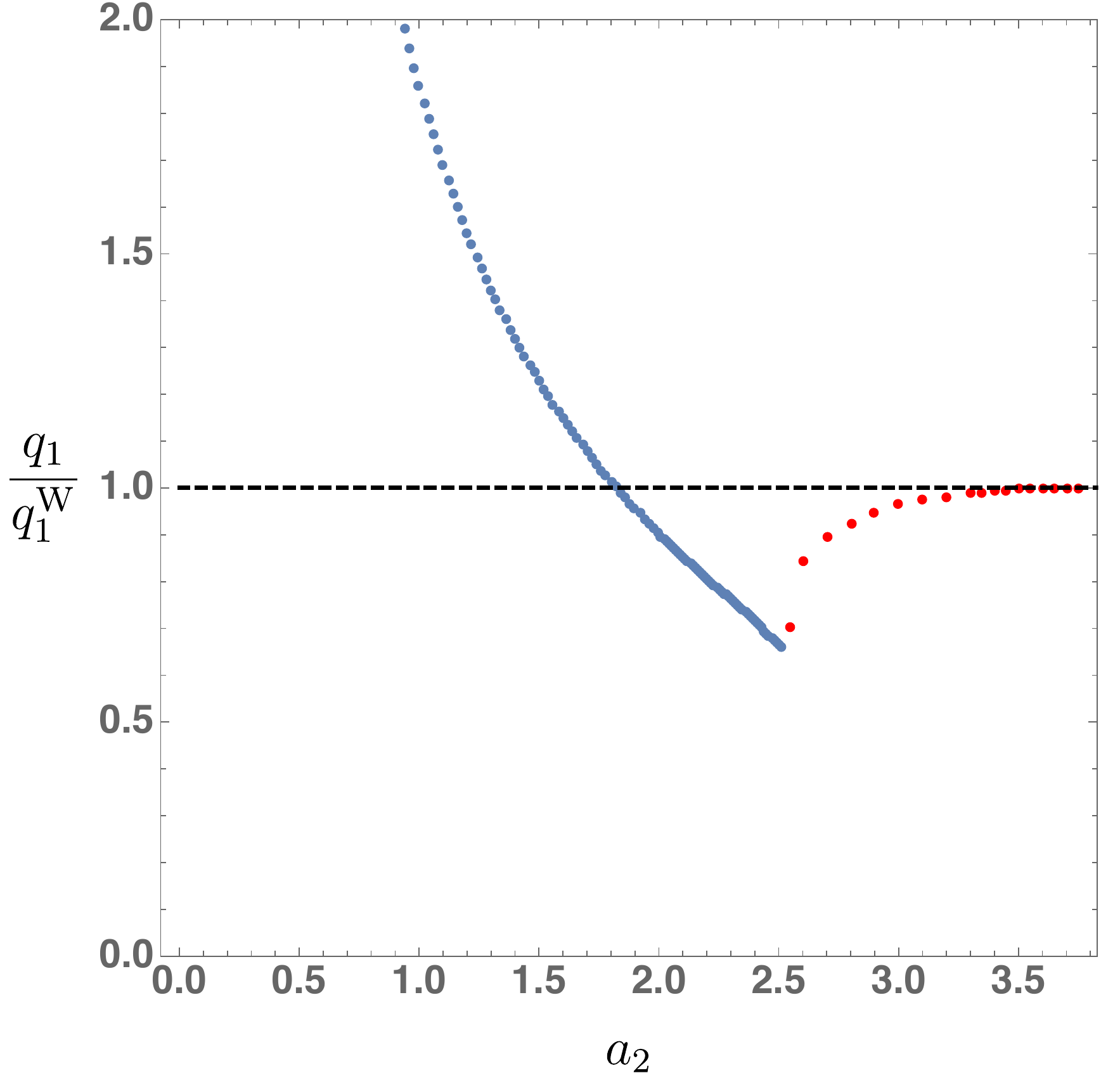}
\caption{Phase diagram of solutions at fixed $a_1=3\,a_2$, $q_2=4$. Solutions with $\Phi_1\neq0$ exist above a line connecting the blue disks and red squares, and $\Phi_1\to0$ along the blue line ($a_2<a_{2\;\max}\approx2.51$), but develops singularities along the red line ($a_2>a_{2\;\max}$). $\Phi_2\ne 0 $ for $a_2$ larger than an amplitude close to $a_{2\;\max}$. }
\label{fig:fixq2}
\end{figure}

So if the two scalar fields satisfy the weak gravity bound, one can no longer violate cosmic censorship. To see if this bound is sharp, we 
need to move the charges below the weak gravity bound and increase the amplitude.
We will consider two cases, each consisting of lowering one of the charges, while keeping the other fixed.  We start by fixing $q_2 = 4$ and lowering $q_1$. The results are shown in Fig.~\ref{fig:fixq2}. At small $a_2$  we plot the minimum charge that is needed to condense $\Phi_1$ around solutions with $\Phi_I=0$ (the blue disks). For larger $a_2$ we follow the condensed phase and lower $q_1$, at fixed $a_2$ and $q_2=4$,  until we find evidence of singular behaviour (the divergence of the Kretschmann scalar). We then mark such points with a red square. Note that the condensed phase  has both $\Phi_I\neq0$ for sufficiently large $a_2$. We find that for large $a_2$, the singular curve approaches the weak gravity bound (we reach it within $0.07\%$). Thus if we try to reduce $q_1$ (at fixed $q_2$) below the weak gravity bound, there is again a maximum amplitude and one could violate cosmic censorship.   This figure is very similar to the results for a single  Maxwell field  \cite{Crisford:2017gsb}.

\begin{figure}[h]
\centering
\includegraphics[width=0.45\linewidth]{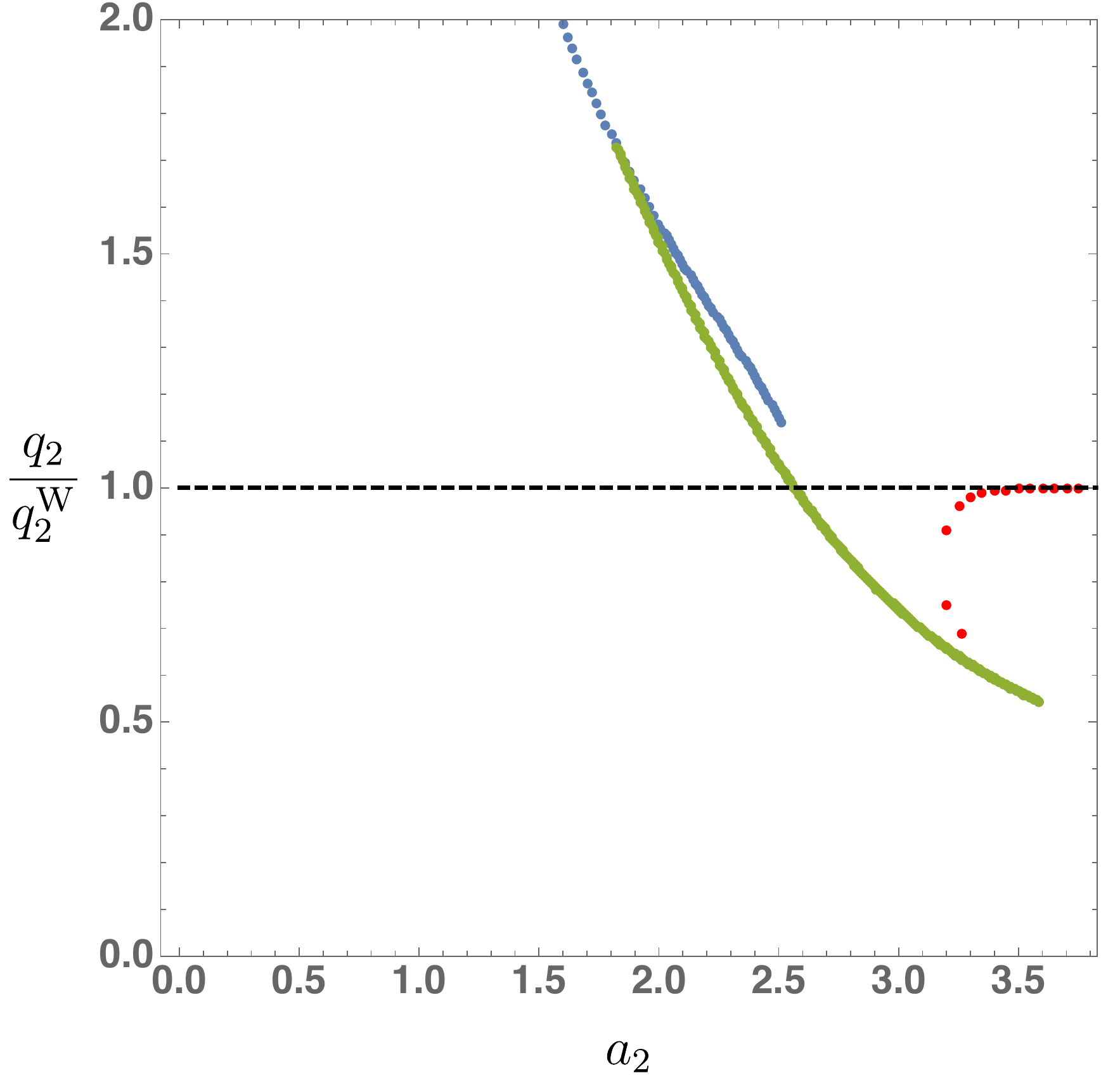}
\caption{Phase diagram of solutions at fixed $a_1=3\,a_2$, $q_1=4/\sqrt{3}$: when increasing $a_2$, the blue disks denote the onset for condensing $\Phi_2$ about the background with no charged scalars, the green diamonds show the onset of $\Phi_2$ about the background with $\Phi_1 \ne 0$, and the red squares represent singular solutions.}
\label{fig:fixq1}
\end{figure}

The results for the second case of fixing $q_1=4/\sqrt{3}$  and lowering $q_2$ are  more surprising and shown in
 Fig.~\ref{fig:fixq1}. The blue disks denote the onset for condensing $\Phi_2$ about the background with no charged scalars. This is analogous to the blue curve in Fig.~\ref{fig:fixq2} but not as relevant after $\Phi_1$ condenses. The green diamonds show the onset of $\Phi_2$ about the background with $\Phi_1 \ne 0$. So above the green curve, both scalars are nonzero.  The red squares represent singular solutions, where the Kretschmann scalar appears to diverge. Notice that the singular points curve around to meet the green curve. Thus there is  a range of $a_2$ where a smooth solution exists  in two disconnected regions of $q_2$. Once again the singular points approach the weak gravity bound for large $a_2$.  (At the last data point, the singularity appears when the charge is just $0.06\%$ lower than $q_2^W$).

Similar results are obtained starting with other choices of $(q_1,q_2)$ that saturate the weak gravity bound (\ref{eq:twofieldbd}). These results show that just like the cases of a single Maxwell field or dilatonic gravity (with $\a <1$),  the condition on the charge to mass ratio of the scalars needed to preserve cosmic censorship for two Maxwell fields is precisely the appropriate weak gravity bound.

\vskip .5in
\centerline {\bf Acknowledgements}
\vskip .2in
It is a pleasure to thank Toby Crisford and Garrett Goon for discussions. GH was supported in part by NSF grant PHY-1801805. JES was supported in part by STFC grants PHY-1504541 and ST/P000681/1.

\vskip .5in
\appendix
\section{The  \emph{Ans\"atze} for numerical solutions\label{sec:secwork}}

To solve the dilatonic field equations (\ref{eqs:EOM}) we use the DeTurck trick, first introduced in \cite{Headrick:2009pv,Adam:2011dn} and reviewed in great detail in \cite{Wiseman:2011by,Dias:2015nua}. We add to Eq.~(\ref{eq:einsteindilaton}) a gauge fixing term
\begin{equation}
R_{ab}+{3}g_{ab}-\nabla_{(a}\xi_{b)}=2\,\nabla_a \phi \nabla_b \phi+2\,e^{-\,2\,\alpha\,\phi}\left(F_{a}^{\phantom{a}c}F_{bc}-\frac{1}{4}g_{ab} F^{cd}F_{cd}\right)\,,
\label{eq:deturckdilaton}
\end{equation}
where $\xi^a=\left[\Gamma^a_{bc}(g)-\Gamma^a_{bc}(\bar{g})\right]g^{bc}$ and $\Gamma^a_{bc}(\mathfrak{g})$ is the Levi-Civitta connection associated with a metric $\mathfrak{g}$. $\bar{g}$ is a reference metric which controls our gauge choice. The DeTurck trick is by now a standard method to find stationary solutions of the Einstein equation, so we will limit ourselves to presenting the reference metric $\bar{g}$, and \emph{Ans\"atze} for $(g,A,\phi)$ and not delve into more details regarding uniqueness of solutions or gauge choice (the interested reader should consult \cite{Figueras:2011va,Dias:2015nua,Figueras:2016nmo}).

We will use the same coordinate system as in \cite{Horowitz:2014gva,Crisford:2017gsb}, which maps standard Poincar\'e coordinates (where we use polar coordinates at the conformal boundary)
\begin{equation}
\mathrm{d}s^2=\frac{1}{z^2}\left(-\mathrm{d}t^2+\mathrm{d}r^2+r^2\,\mathrm{d}\varphi^2+\mathrm{d}z^2\right)\,,
\end{equation}
into
\begin{equation}
\mathrm{d}s^2=\frac{1}{(1-x^2)^2}\left[-\frac{(1-y^2)^2\,\mathrm{d}t^2}{y^2(2-y^2)}+\frac{4\mathrm{d}y^2}{y^2(1-y^2)^2(2-y^2)^2}+\frac{4\mathrm{d}x^2}{2-x^2}+x^2(2-x^2)\mathrm{d}\varphi^2\right]
\label{eq:pureads}
\end{equation}
via the map
\begin{subequations}
\begin{align}
&z=\frac{y\sqrt{2-y^2}}{1-y^2}(1-x^2)\,,
\\
&r=\frac{y\sqrt{2-y^2}}{1-y^2}x\sqrt{2-x^2}\,.
\end{align}
\end{subequations}
In the $(x,y)$ coordinates, $y=0$ corresponds to the boundary point $r=z=0$ where the boundary intersects the axis of rotation, $x=1$ is the location conformal boundary, $y=1$ the Poincar\'e horizon and $x=0$ the axis of rotation.

At the boundary ($x=1$), we require a gauge potential that has only a nonzero time component:
\begin{equation}
A_{\partial}= \frac{a}{\displaystyle \left(1+{r^2}\right)^{4}}\,\mathrm{d}t\,.
\end{equation}
 At the conformal boundary, $\left.r\right|_{x=1}=y\sqrt{2-y^2}/(1-y^2)$, which gives
\begin{equation}
A_{\partial}= a(1-y^2)^{8}\,\mathrm{d}t\,.
\end{equation}

The dilaton has zero mass and, using standard quantisation, this means that we expect that $\phi \sim z^3\sim (1-x)^3$ as we approach the conformal boundary. Numerically, dealing with exponentials is not an easy to task, so we perform the following function redefinition
\begin{equation}
\phi=\frac{1}{\alpha}\mathrm{arctanh}\left(\alpha\, \hat{\phi}\right)\,.
\end{equation}
Note that, at the boundary, $\hat{\phi}\sim \phi$.

We are now ready to present our \emph{Ans\"atze} for all fields:
\begin{subequations}
\begin{align}
&\mathrm{d}s^2=\frac{1}{(1-x^2)^2}\Bigg\{-\frac{(1-y^2)^2\,Q_1(x,y)\,\mathrm{d}t^2}{y^2(2-y^2)}+\frac{4\,Q_2(x,y)\,\mathrm{d}y^2}{y^2(1-y^2)^2(2-y^2)^2}\nonumber
\\
&\qquad\qquad\qquad\qquad +\frac{4\,Q_4(x,y)}{2-x^2}\left[\mathrm{d}x+\frac{Q_3(x,y)}{1-y^2}\mathrm{d}y\right]^2+x^2(2-x^2)\,Q_5(x,y)\,\mathrm{d}\varphi^2\Bigg\}\,,\label{eq:metric}
\\
&A =  Q_6(x,y)\, \mathrm{d}t\,,
\\
&\phi=\frac{1}{\alpha}\mathrm{arctanh}\left[\alpha\,(1-x^2)^3\,Q_7(x,y)\right]\,,
\end{align}
\end{subequations}
where $Q_I(x,y)$,  $I\in\{1,\ldots,7\}$ are functions of $(x,y)$ to be determined in our numerical procedure. For the reference metric, we take $Q_1=Q_2=Q_4=Q_5=1$ and $Q_3=0$, corresponding to Eq.~(\ref{eq:pureads}).

We next discuss our boundary conditions. At the conformal boundary we demand $Q_1(1,y)=Q_2(1,y)=Q_4(1,y)=Q_5(1,y)=1$, $Q_3(1,y)=0$ and $Q_6(1,y)=a\,(1-y^2)^n$, while for $Q_7$ we find a simple Robin boundary condition by solving the Einstein-DeTurck PDE system (\ref{eq:deturckdilaton}) order by order in $(1-x)$:
\begin{equation}
(1-y^2)^2 \left.\frac{\partial Q_7}{\partial x}\right|_{x=1}+\frac{\alpha}{8}y^2(2-y^2)\left[\left(\left.\frac{\partial Q_6}{\partial x}\right|_{x=1}\right)^2+y^2(2-y^2)^2(1-y^2)^2\left(\left.\frac{\partial Q_6}{\partial y}\right|_{x=1}\right)^2\right]=0\,.
\end{equation}

At the Poincar\'e horizon, located at $y=1$, we demand $Q_1=Q_2=Q_4=Q_5=1$ and $Q_3=Q_6=Q_7=0$. At the axis of rotation, located at $x=0$, we impose
\begin{align}
&\left.\frac{\partial Q_1}{\partial x}\right|_{x=0}=\left.\frac{\partial Q_2}{\partial x}\right|_{x=0}=\left.\frac{\partial Q_4}{\partial x}\right|_{x=0}=\left.\frac{\partial Q_5}{\partial x}\right|_{x=0}=\left.\frac{\partial Q_6}{\partial x}\right|_{x=0}=\left.\frac{\partial Q_7}{\partial x}\right|_{x=0}=0\nonumber
\\
&Q_3(0,y)=Q_4(0,y)-Q_5(0,y)=0\,,
\end{align}
with the latter condition enforcing a $2\pi$ period for $\varphi$. Finally, at $y$=0 we demand
\begin{align}
&Q_1(x,0)=Q_2(x,0)=Q_4(x,0)=Q_5(x,0)=1\,,\quad Q_3(x,0)=0\,,\nonumber
\\
&Q_6(x,0)=a\,,\quad\text{and}\quad Q_7(x,0)=0\,.
\end{align}

Next we discuss the  boundary conditions for the charged scalar $\Phi$. Since $\Phi$ has conformal dimension $\Delta$ we expect $\Phi \sim z^\Delta$ close to $z\sim0$, that is to say that $\Phi$ vanishes as $(1-x)^\Delta$ close to $x=1$. Similarly close to $y=0$, we expect $\Phi$ to vanish as $y^\Delta$, as such we perform a function redefinition of the form
\begin{equation}
\Phi = (1-x^2)^\Delta y^\Delta (2-y^2)^{\Delta/2}\,Q_8\,.
\end{equation}
We adopt this function redefinition both at the linear and nonlinear level. At the boundary, axis of symmetry and $y=0$ we find pure Neumann boundary condition
\begin{equation}
\left.\frac{\partial Q_8}{\partial x}\right|_{x=0}=\left.\frac{\partial Q_8}{\partial x}\right|_{x=1}=\left.\frac{\partial Q_8}{\partial y}\right|_{y=0}=0\,,
\end{equation}
while at $y=1$, the Poincar\'e horizon we impose a Dirichlet condition
\begin{equation}
Q_8(x,1)=0\,.
\end{equation}
These boundary conditions hold both at the linear and nonlinear level.

In Fig.~\ref{fig:largecond}, we plot the operator $\mathcal{O}$ dual to $\Phi$ evaluated at the origin on the boundary. This is simply given by
\begin{equation}
\langle \mathcal{O} \rangle = (1-y^2)^2Q_8(1,y)\Rightarrow \left.\langle \mathcal{O}\rangle\right|_{r=0} = Q_8(1,0)\,.
\end{equation}

To check our claim that there are no smooth solutions for $a>a_{\max}$ without the charged scalar, or for $q<q_{\min}$ with a charged scalar, we
 also tried to use DeTurck flow \cite{Figueras:2011va}. However, we did not find any regular solution. Instead, the system approached a singular solution at late flow times.

To find the nonlinear solutions with two Maxwell fields and two charged scalars numerically, most of the work regarding our choice of boundary conditions and \emph{Ans\"atze} for the several fields is the same as above. For the metric \emph{ansatz} we choose a line element given by (\ref{eq:metric}), while for the gauge and scalar fields we take
\begin{equation}
A_I =  Q_{5+I}(x,y)\,,\quad\text{and}\quad\Phi_{I} = y^2(2-y^2)(1-x^2)^2Q_{7+I}(x,y)\,.
\end{equation}
The boundary conditions are the same as above.

\bibliography{further_evidence}{}
\bibliographystyle{JHEP}
\end{document}